\documentclass[11pt,dvips]{article}
\usepackage{epsfig,times} 

\usepackage{picinpar}
\usepackage{wrapfig}
\usepackage{floatflt}
\makeatletter
\renewcommand{\section}{\@startsection{section}{1}{0in}
	{0.4\baselineskip}{0.1\baselineskip}{\Large\bf}}
\renewcommand{\subsection}{\@startsection{subsection}{2}{0in}
	{0.25\baselineskip}{-\baselineskip}{\large\bf}}
\renewcommand{\subsubsection}{\@startsection{subsubsection}{3}{0in}
	{0.1\baselineskip}{-\baselineskip}{\normalsize\bf}}
\makeatother
\begin{document}

%
%  Session and Paper Code:
\thispagestyle{myheadings}
%
%  ***INSTRUCTIONS:***  Replace `OG 9.9.9' in the command argument below
%			with your assigned session and paper code:
\markright{HE 3.1.1}
\begin{center}
%
%  ***INSTRUCTIONS:***  Replace `Instructions for Preparation of Manuscript'
%			with your paper's title:
{\LARGE \bf Cygnus X-3 Revisited: 10 Years of Muon and Radio Observations}
\end{center}

%  Author List:
\begin{center}
%
%  ***INSTRUCTIONS:***  Replace authors and addresses below with your own:
%
{\bf W.W.M.~Allison,$^3$, G.J.~Alner$^4$, D.S.~Ayres$^1$,
W.L.~Barrett,$^6$, C.~Bode,$^2$, P.M.~Border$^2$, C.B.~Brooks$^3$, 
J.H.~Cobb$^3$, R.J.~Cotton$^4$,
H.~Courant$^2$, D.M.~Demuth$^2$, T.H.~Fields$^1$, H.R.~Gallagher$^3$, 
M.C.~Goodman$^1$, R.~Gran$^2$, T.~Joffe-Minor$^1$, T.~Kafka$^5$, S.M.S.~Kasahara$^2$,
W.~Leeson$^5$, P.J.~Litchfield$^4$, N.P.~Longley$^2$, W.A.~Mann$^5$, 
M.L.~Marshak$^2$, R.H.~Milburn$^5$,
W.H.~Miller$^2$, L.~Mualem$^2$, A.~Napier$^5$, W.P.~Oliver$^5$, G.F.~Pearce$^4$,
E.A.~Peterson$^2$, D.A.~Petyt$^4$, L.E.~Price$^1$, K.~Ruddick$^2$, M.~Sanchez$^5$,
J.~Schneps$^5$, M.H.~Schub$^2$, R.~Seidlein$^1$, A.~Stassinakis$^3$, J.L.~Thron$^1$,
V.~Vassiliev$^2$, G.~Villaume$^2$, S.~Wakely$^2$, N.~West$^3$, D.~Wall$^5$\\
(The Soudan 2 Collaboration)}\\
{\it $^{1}$Argonne National Laboratory, Argonne, IL 60439, USA\\
$^{2}$University of Minnesota, Minneapolis, MN 55455, USA\\
$^3$Department of Physics, University of Oxford, Oxford OX1 3RH, UK\\
$^4$Rutherford Appleton Laboratory, Chilton, Didcot, Oxfordshire OX11 0QX, UK\\
$^5$Tufts University, Medford MA 02155, USA\\
$^6$Western Washington University, Bellingham, WA 98225, USA}
\end{center}

%  Abstract:
\begin{center}
{\large \bf Abstract\\}
\end{center}
\vspace{-0.5ex}

The Soudan 2 deep underground tracking calorimeter has recorded cosmic ray
muon tracks from the direction of the galactic x-ray binary Cygnus X-3
on most transits during the interval 1989-1998. We analyze these events in the
context of previous reports of Cygnus X-3-related muon flux during major radio
flares of that source. We find some evidence for excess flux during a small
number of transits coincident with major radio flares. We also find an
indication that these events may be distributed around the source with a Gaussian
point spread function with $\sigma = 1.3^{\circ}$, larger than the instrumental angular spread
of $\le 0.3^{\circ}$, verified by observation of the shadow of the moon. 

%

%  Leave this line skip in place:
\vspace{1ex}

\section{Introduction}
\label{intro.sec}

Cygnus X-3 is a galactic binary star system known to emit highly
variable fluxes of radio, infrared and x-ray radiation (Bonnet-Bidaud, 1988
and Nagle, 1988). Episodically, radio fluxes from Cygnus X-3 vary over three orders of
magnitude within a few days (Waltman, 1994). However, such major flares are
infrequent, occurring on average every several years.
The best current understanding of Cygnus X-3 is that it is one of several known
"microquasars," non-thermal star systems in our galaxy that from time to time
emit strong radio flares associated with relativistic jets pointed towards the earth.
Microquasars are smaller versions of extragalactic quasars, which also produce highly
collimated radiation through relativistic jets.

Over a number of years, cosmic ray air shower and deep underground
muon detectors have reported both positive and negative observations 
of TeV or above quanta associated with Cygnus X-3. Several observations 
have identified high energy muons as secondaries in Cygnus X-3-related
events. At TeV energies, only stable, neutral particles can
travel the $>8$ kpc distance from Cygnus X-3 to the earth along trajectories
which point back to the source. The known stable, neutral particles--photons
and neutrinos--have only small probabilities for producing detectable
muons. Thus, observation of TeV muons associated with Cygnus X-3 requires
either exotic interactions of known primaries, exotic primaries or very
large fluxes of neutrinos or photons. The lack of a conventional physical model and     
the sometimes contradictory reports of transient fluxes have decreased
recent interest in Cygnus X-3 as a TeV or higher energy source. For the past 5 years,
there have been no published reports concerning such particles from Cygnus X-3.

The last published, positive Cygnus X-3 muon observation reported excess deep 
underground muons observed by the Soudan 2 detector 
(Thomson, 1991: Paper I and Thomson, 1992) apparently in association 
with that system's January 1991 major radio flare. Other nearly
simultaneous observations observed no unexpected effects (Becker-Szendy, 1993).
There are several reasons to now re-examine the Soudan 2 observations of muons
from the direction of Cygnus X-3. (1) Paper I was based on
2.1 years of data, during an interval when the Soudan 2 detector
was under construction. Soudan 2 has now recorded 10 years of data during
mostly stable operation. (2) The analysis for Paper I used a conservative estimate
of the angular resolution and pointing accuracy of the Soudan 2 detector. These
parameters have now been measured by observation of the
cosmic ray shadow of the moon. The actual values (and 1991 estimates) for
angular resolution and pointing accuracy are: gaussian with $\sigma \le 0.3^{\circ}
~(1.0^{\circ})$ and $\le0.15^{\circ}~(0.5^{\circ})$. (3) Detailed radio
flux measurements for Cygnus X-3 are now available for most days during the entire 10-year
Soudan 2 observation interval.

\section{Data Collection and Analysis}
\label{data.sec}

The Soudan 2 detector is a deep underground iron tracking calorimeter
designed to search for proton decay. The detector and the event collection
and track analysis procedures are described elsewhere (Allison, 1996). The data sample
reported here consists of $3.6 \times 10^7$ events which have passed run
and event cuts, both intended to identify a high-purity, high-resolution
muon track sample. The backgrounds or expected numbers of events in the absence
of a source reported here have been determined by randomly pairing, month-by-month,
arrival times of real events with track directions in detector coordinates of
other real events. This analysis uses a $100 \times$ background sample.

As in Paper I, we focus first on real and background events which point within $2^{\circ}$
of the direction of Cygnus X-3 (J2000 $\alpha = 308.10^{\circ}, \delta = 40.96^{\circ}$).
From the radio data, we have identified 3 major and 7 intermediate radio flares of
Cygnus X-3 during the entire 1989-1998 interval as listed in the table below.
The entire data sample includes 3,046 Cygnus X-3 transits during which the background calculation
indicates that 2 or more events are expected from the source direction. Of these
transits, 139 are coincident with major radio flares, 105 with intermediate radio
flares and 2,802 with neither of these categories of radio flares. For each transit,
we have determined the Poisson probability $P(\ge n,\mu)$ that
$n$ or more muons are observed (where n is the number of muons that were
observed) during a transit in which $\mu$ muons are expected. 
Fig. 1(a) shows a semi-log histogram
of the number of transits $N$ vs. $-\log_{10} P(\ge n,\mu)$ for the major flare,
intermediate flare and neither flare transits. The figure also shows Monte Carlo calculations
of the same distributions, assuming Poisson statistics.  

\vspace{1ex}
Table 1. Major and Intermediate Radio Flares of Cygnus X-3: 1989 to 1998.
\vspace{0.1in}

\begin{tabular}{|ccccc|}
\hline
Type&Start Date&End Date&Peak 3 GHz Flux (Jy)&Peak 8 GHz Flux (Jy)\\
\hline
Major&1 Jun 89&14 Aug 89&15.4&15.8\\
Intermediate&11 Aug 90&31 Aug 90&7.8&7.2\\
Intermediate&1 Oct 90&22 Oct 90&9.3&9.0\\
Major&18 Jan 91&19 Mar 91&11.9&13.9\\
Intermediate&19 Jun 91&30 Jun 91&4.5&3.6\\
Major&24 Jul 91&29 Aug 91&16.4&13.1\\
Intermediate&1 Sep 92&11 Sep 92&3.7&3.7\\
Intermediate&20 Feb 94&11 Mar 94&4.8&4.6\\
Intermediate&1 Feb 97&16 Feb 97&9.2&8.4\\
Intermediate&10 Jun 97&25 Jun 97&3.7&3.1\\
\hline
\end{tabular}
\vspace{1ex}

The distributions in Fig. 1(a) for the ``Intermediate Flare''
and ``No Flare'' transits appear consistent with expectations of
Poisson statistics. However, as suggested in Paper I, there are transits in the ``Major
Flare'' sample with an excess of observed events, including the two specific
transits cited in that paper. For example, the "Major Flare" graph in Fig. 1(a)
shows 6 transits with $-\log_{10} P(\ge n,\mu) \ge 2$ with 0.89 such transits expected
(chance probability $\le 3.2 \times 10^{-4}$) and 10 transits with 
$-\log_{10} P(\ge n,\mu) \ge 1.5$ with 3.1 such transits expected
(chance probability $\le 1.4 \times 10^{-3}$). Of these 10 transits,
3 occurred during the June 1989 flare, 4 (including the two cited in Paper 1)
occurred during the January 1991 flare and 3 occurred during the August 1991 flare
(see Table 2). The numbers of events listed in the table for the transits discussed in
Paper I differ somewhat from those given in that paper because the current
analysis uses a modified reconstruction algorithm and tighter cuts
in order to improve the detector angular resolution and reduce reconstruction
errors.  

\vspace{1ex}
Table 2. Transits with $-\log_{10} P(\ge n,\mu) \ge 1.5$.
\vspace{0.1in}

\begin{tabular}{|cccc|}
\hline
Date&Observed Events&Expected Events&$-\log_{10} P(\ge n,\mu)$\\
\hline
5 Jun 89&9&2.53&2.91\\
18 Aug 91&17&7.62&2.64\\
24 Aug 91&16&7.45&2.36\\
24 Jan 91&13&5.62&2.28\\
16 Aug 91&15&7.15&2.16\\
19 Jun 89&8&2.79&2.10\\
10 Mar 91&12&5.61&1.89\\
19 Feb 91&11&5.16&1.77\\
15 Jun 89&7&2.75&1.65\\
21 Jan 91&10&4.97&1.51\\
\hline
\end{tabular}
\vspace{1ex}

Although a $2^{\circ}$ half-angle cone has been used to select the transits
discussed above, the $\le0.3^{\circ}$ angular resolution of the Soudan 2
detector should indicate pointing towards the source, if an event excess
associated with major radio flares is
a real effect. Fig. 1(b) shows an angular distribution for all 4,303 muon events
observed within $5^{\circ}$ of Cygnus X-3 during
all transits occurring during the three major radio flares 
(not just the 10 transits listed in Table 2). 
A goodness-of-fit test between the observed distribution in Fig. 1(b) and the
expected (or background) distribution in the absence of a source (not shown) yields
$\chi^2 = 37.4$ for 25 df (chance probability $= 0.05$). The fit shown in
Fig. 1(b) adds a Gaussian source (or sink) term to the expected ``no-source''
distribution with the amplitude of the Gaussian fixed by the total number of
excess observed events. A $\chi^2$ minimization then yields $\sigma = 1.3^{\circ}  
\mbox{}^{+0.5^{\circ}}_{-0.4^{\circ}}$. The chance probability of the $\chi^2$
improvement resulting from  the fit is 0.01. The data in Fig. 1(b) show no evidence
for a source with the detector angular resolution of $\sigma \le 0.3^{\circ}$.
Also, plots similar to Fig. 1(b) for data recorded during the ``Intermediate Flare''
and ``No Flare'' intervals show no evidence for any source term in their
angular distributions.

\section{Conclusions}
\label{conclusions.sec}

The question of whether VHE or UHE quanta are associated with Cygnus X-3 may
remain unresolved for some period of time. We have shown
here evidence for an excess of directional events associated with
major radio flares of Cygnus X-3. However, the absence of a well-understood
physical mechanism, the ${\it post~hoc}$ nature of the analysis and the
small number of events leading to modest statistical confidence levels
diminish the impact of these observations. What is required to solidify
any conclusions about high energy quanta from Cygnus X-3 are predictable observations. 
Yet, despite daily monitoring, no major Cygnus X-3 radio flares 
have been observed since 1991, making predictions about muons
during major radio flares untestable. It is not known if or when
such flares might be observed in the future. Thus, searches for deep underground
muon sources may need to be directed elsewhere, if past observations are ever to be
understood.

The authors thank Dr. E. Waltman for providing access to Cygnus X-3 radio monitoring
data in machine-readable format.

%
%  References: (DO NOT ALTER NEXT 4 LINES)
\vspace{1ex}
\begin{center}
{\Large\bf References}
\end{center}
%
%  ***INSTRUCTIONS:***  Enter your references alphabetically following the format
%			of the example citations below.
Allison,~W.W.M.,~{\it et al.}, 1996 Nucl. Inst. and Meth. \textbf {A376},
377 and \textbf {A381}, 385.\\
Becker-Szendy,~R., {\it et al.}, Phys. Rev. {\bf D10}, 4203 (1993).\\
Bonnet-Bidaud,~J.M. and G.~Chardin, 1988 Phys. Rep. {\bf 170}, 326.\\
Nagle,~D.E., T.K.~Gaisser and R.J.~Protheroe, 1988 Ann. Rev. Nucl. Part. Sci., 609.\\
Waltman,~E.B., {\it et al.}, 1994 Astronom. J. {\bf 112}, 2690 (1996), {\bf 110}, 290 (1995),
{\bf 108}, 179.\\
Thomson,~M.A., {\it et al.}, Phys. Lett. {\bf B269}, 220 (1991).\\
Thomson,~M.A., D. Phil. Thesis, Oxford University, 1992.\\

\begin{figwindow}[1,r,%
{\mbox{\epsfig{file=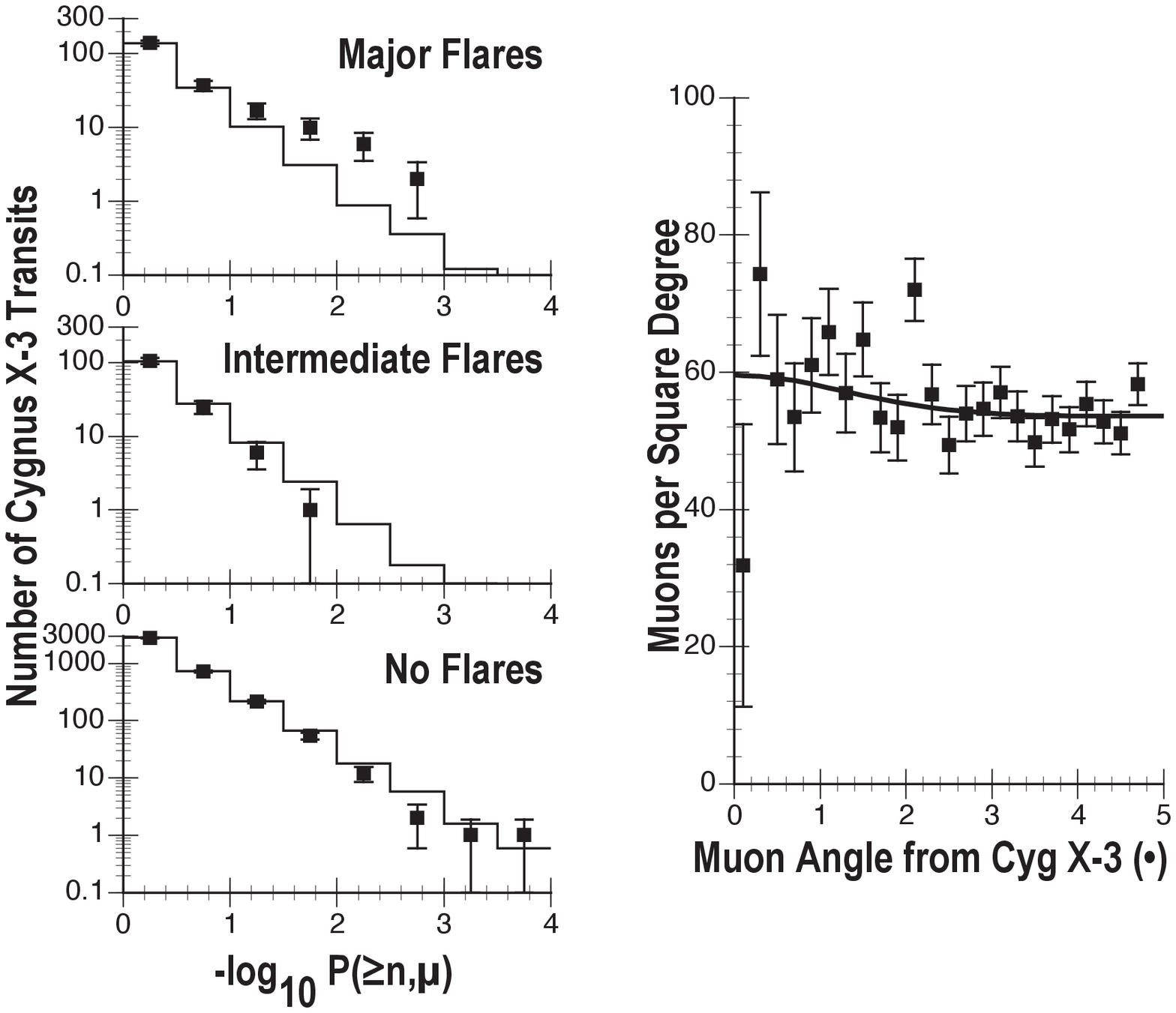,width=6.6in}}},%
{(a) (left) The observed numbers of transits (points) vs.
$-\log_{10} P(\ge n,\mu)$ for ``Major Flare'', ``Intermediate Flare'' and
``No Flare'' intervals described in the text. The solid lines show the
expected distributions from a simulation that assumes Poisson statistics. 
(b) (right) The number of muons per square degree vs. the angle in degrees
between the muon track and Cygnus X-3 for events during major radio flares.
The solid line is the Gaussian fit described in the text with $\sigma = 1.3^{\circ}$. }]
\end{figwindow}

\end{document}